\documentclass[sigconf]{acmart}

\AtBeginDocument{%
  \providecommand\BibTeX{{%
    \normalfont B\kern-0.5em{\scshape i\kern-0.25em b}\kern-0.8em\TeX}}}

\setcopyright{acmlicensed}
\copyrightyear{2018}
\acmYear{2018}
\acmDOI{XXXXXXX.XXXXXXX}

\acmConference[Conference acronym 'XX]{Make sure to enter the correct
  conference title from your rights confirmation emai}{June 03--05,
  2018}{Woodstock, NY}
\acmISBN{978-1-4503-XXXX-X/18/06}

\usepackage{multirow}
\usepackage{multicol}
\usepackage{listings}
\usepackage{makecell}
\usepackage{array}
\usepackage{float}
\usepackage{microtype}
\usepackage{graphicx}
\usepackage{subfigure}
\usepackage{booktabs} 
\usepackage{enumitem}
\usepackage{wrapfig}
\usepackage{tcolorbox}
\usepackage{colortbl} 
\usepackage{xcolor} 

\usepackage{amsmath} 
\usepackage{caption} 
\usepackage{graphicx}
\usepackage[ruled,vlined]{algorithm2e}

\theoremstyle{definition}

\theoremstyle{plain}

\renewcommand{\shortauthors}{Trovato and Tobin, et al.}

\title{Solving the Content Gap in Roblox Game Recommendations: LLM-Based Profile Generation and Reranking}

\author{Chen~Wang}
\authornote{This author worked as an intern at Roblox.}
\affiliation{%
  \institution{University of Illinois Chicago}
  \city{Chicago}
  \state{Illinois}
  \country{USA}
}
\email{cwang266@uic.edu}

\author{Xiaokai~Wei, Yexi~Jiang, Frank~Ong, Kevin~Gao, Xiao~Yu}
\affiliation{%
  \institution{Roblox}
  \city{San Mateo}
  \state{California}
  \country{USA}
}
\email{{xwei, hjiang, fong, kgao, xyu}@roblox.com}

\author{Zheng~Hui}
\affiliation{%
  \institution{Columbia University}
  \city{New York}
  \state{New York}
  \country{USA}
}
\email{zh2483@columbia.edu}

\author{Se-eun~Yoon}
\affiliation{%
  \institution{University of California, San Diego}
  \city{La Jolla}
  \state{California}
  \country{USA}
}
\email{seeuny@ucsd.edu}

\author{Philip~Yu}
\affiliation{%
  \institution{University of Illinois Chicago}
  \city{Chicago}
  \state{Illinois}
  \country{USA}
}
\email{psyu@uic.edu}

\author{Michelle~Gong}
\affiliation{%
  \institution{Roblox}
  \city{San Mateo}
  \state{California}
  \country{USA}
}
\email{mgong@roblox.com}

\renewcommand{\shortauthors}{Chen Wang et al.}

\begin{abstract}
With the vast and dynamic user-generated content on Roblox, creating effective game recommendations requires a deep understanding of game content. However, due to the inconsistent and sparse nature of game text features—such as titles and descriptions—traditional recommendation models struggle to interpret and utilize this information effectively. Recent advancements in large language models (LLMs) offer new possibilities for enhancing recommendation systems by extracting and analyzing in-game text data. This paper addresses two primary challenges: \textbf{(1) generating high-quality, structured text features for games without extensive human annotation}, and \textbf{(2) validating the quality of these features to ensure they improve recommendation relevance}. To tackle these, we propose an approach centered on extracting in-game text and leveraging LLMs to infer essential attributes, such as genre and gameplay objectives, from raw player interactions. Additionally, we introduce an LLM-based re-ranking mechanism to verify the effectiveness of the generated text features, thereby enhancing personalization and user satisfaction. Beyond recommendations, our approach enables applications such as user engagement-based integrity detection, which has already been deployed in production. Our method provides a scalable, content-driven framework for improving recommendation quality on Roblox, demonstrating the potential of in-game text understanding to adapt recommendations to the platform's unique, user-generated ecosystem.
\end{abstract}

\begin{document}

\begin{teaserfigure}
\centering
\includegraphics[width=\linewidth]{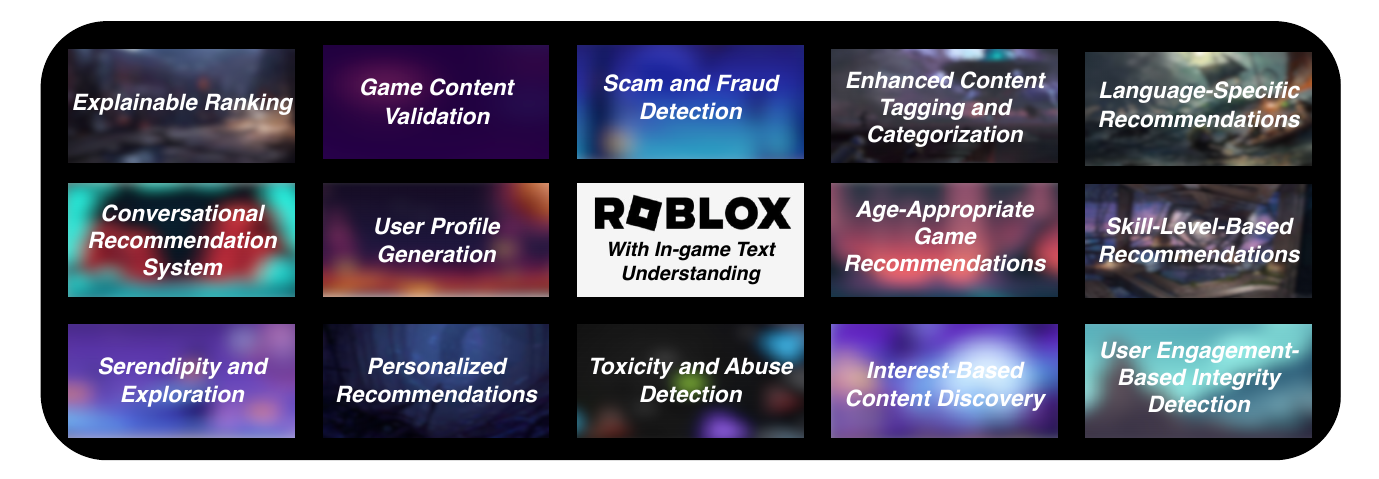}
\caption{Potential Applications of In-Game Text Understanding on Roblox. This figure demonstrates the broad range of applications enabled by in-game text understanding on the Roblox platform. By leveraging insights from game content, Roblox can enhance user experience, improve recommendation accuracy, and strengthen content integrity, supporting a more personalized, engaging, and safe environment for its players.}
\label{fig:roblox_logo}
\end{teaserfigure}



\maketitle

\newcommand{\checkvalue}[2]{\ifdim #1pt > #2pt \textcolor{blue}{#1}\else \textcolor{red}{#1}\fi}

\section{Introduction}

Roblox is a popular online platform where users create and play games designed by other users, resulting in a vast and diverse collection of interactive experiences. To enhance user engagement, Roblox relies on a multi-stage recommendation system that ranks games based on deep neural network (DNN)-driven models, leveraging sparse ID features (such as User ID, Game ID) and dense features derived from user behavior and game statistics. However, these recommendations often fail to capture individual user preferences fully, as they lack content-based signals—particularly game-related text features such as genre and descriptions.

Recent advances in large language models (LLMs) offer new opportunities to improve recommendation systems by leveraging game text data. LLMs can deepen the understanding of game content, enabling not only enhanced personalization but also supporting essential tasks such as game content validation, scam and fraud detection, and age-appropriate recommendations. As illustrated in Figure~\ref{fig:roblox_logo}, in-game text understanding enables a broad range of applications, including explainable ranking, toxicity detection, and conversational recommendations, fostering user trust by providing transparent and engaging interactions. 

Re-ranking techniques refine recommendation lists to better align with user preferences, balancing factors like accuracy, diversity, and personalization~\cite{gao2024llm, carraro2024enhancing}. Recent advancements leverage LLMs, including Chain-of-Thought (CoT) reasoning~\cite{gao2024llm} and instruction-tuning frameworks like RecRanker~\cite{luo2023recranker}, which use enriched prompts to enhance personalization. Transformer-based models further improve quality by modeling item relationships and user preferences holistically~\cite{gao2024llmenhancedrerankingrecommendersystems}. While effective in structured environments, these methods are underexplored in dynamic, unstructured platforms like Roblox. Our work extends these approaches, adapting LLMs to handle Roblox’s noisy game text data, enabling personalized re-ranking in a user-generated ecosystem.

Unlike platforms like Steam~\cite{yang2022large,cheuque2019recommender,pathak2017generating}, where game text is structured and professionally crafted, Roblox faces unique challenges. With accessible tools like Roblox Studio, even young users can create games, resulting in inconsistent and unstructured game text, such as titles and descriptions. This variability, coupled with the rapid influx of new games, complicates the use of text-based features for recommendations. While LLMs excel with structured text, they struggle with Roblox’s unrefined descriptions. Generating high-quality, reliable text features without extensive human annotation is essential for delivering personalized recommendations.

This paper addresses two critical challenges in improving game recommendations on Roblox, where the platform’s user-generated content results in inconsistent and sparse game text features, such as titles and descriptions. This scenario creates a “chicken-and-egg” problem: effective game recommendations require high-quality text features, but without structured or professionally curated descriptions, LLMs struggle to interpret game content—particularly for new or rapidly changing games.

The first challenge is to \textbf{develop a method for generating high-quality, structured text features for Roblox games without extensive human annotation}, which is infeasible given Roblox’s scale and rapid content turnover. The second challenge is to \textbf{establish a framework to validate the quality of these generated text features} to ensure that they enhance recommendation accuracy. Addressing these challenges is essential for building a scalable, content-driven recommendation system that adapts to Roblox's unique dynamics.

To address the first challenge—generating high-quality game text features—we propose a method centered on extracting and understanding raw in-game text. Our approach leverages the fact that developers often guide players with in-game instructions to prevent drop-off. Using LLMs equipped with strong global knowledge, we analyze this in-game text to infer attributes such as game genre, content themes, and player objectives, ultimately constructing a high-quality, structured game profile. To tackle the second challenge—validating the quality of these generated profiles—we introduce an LLM-based re-ranking mechanism. This model integrates the generated text features to validate ranking performance and personalize game recommendations. By re-ranking based on text feature quality, we ensure that our generated profiles effectively improve both recommendation relevance and user satisfaction.

In summary, the key contributions of this paper are as follows:
\begin{itemize}[leftmargin=*]
\item \textbf{In-Game Text Extraction and Understanding}: We propose a novel approach for generating high-quality game profiles on Roblox by extracting and interpreting raw in-game text, using LLMs to infer game attributes such as genre, content, and play style without human annotation.
\item \textbf{LLM-Based Re-Ranking for Quality Verification}: To validate the effectiveness of generated game profiles, we introduce an LLM-based re-ranking model that incorporates text features into the recommendation system, enhancing ranking personalization and relevance.
\item \textbf{Scalable Framework for Content-Driven Recommendations}: By addressing the challenges of text feature generation and quality validation, this work lays the foundation for scalable, content-based recommendation improvements on Roblox, adaptable to the platform’s dynamic, user-generated ecosystem.
\end{itemize}
\section{Related Works}

Content-based recommendation systems enhance personalization by leveraging textual and content-derived features. Early methods relied on structured metadata, such as product descriptions and user reviews~\cite{degemmis2007content,aciar2007informed}, while recent advances in deep learning extract rich representations from unstructured text~\cite{liu2024once,bondevik2024systematic}. In gaming, studies have explored large-scale game recommendations~\cite{yang2022large}, user preferences in online games~\cite{POLITOWSKI2018103}, and action recommendations in text-based games~\cite{Recommend_Actions_in_Text_Games}. However, these approaches often assume clean, structured input data, which is unavailable on user-generated platforms like Roblox, where text is frequently noisy and inconsistent. Our work addresses this gap by extracting raw in-game text and generating structured profiles to capture genre, objectives, and gameplay mechanics, enabling effective recommendations in noisy environments.
\begin{figure}[t]
\begin{center}
\centerline{\includegraphics[width=\linewidth]{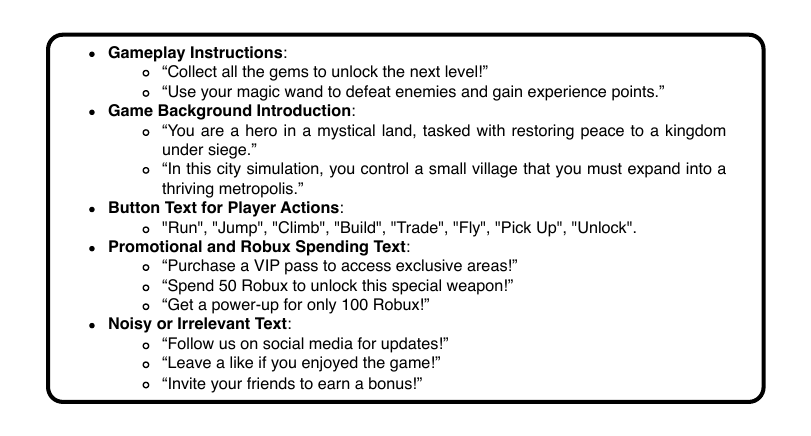}}

\caption{Illustration of the various types of in-game text in Roblox games, including: (1) Gameplay Instructions, guiding player actions and objectives; (2) Game Background Introduction, providing insights into genre and narrative; (3) Button Text for Player Actions, indicating mechanics like movement and interaction; (4) Promotional and Robux Spending Text; and (5) Noisy or Irrelevant Text.}
\label{fig:in_game_text}
\end{center}
\vskip -0.1in
\end{figure}
\begin{algorithm}[htbp]
\caption{Game Profile Generation and LLM-Based Reranking}
\label{alg:reranking}
\DontPrintSemicolon
\SetKwInOut{Input}{Input}\SetKwInOut{Output}{Output}

\Input{
- Set of games $\mathcal{G}$

- For each game $g \in \mathcal{G}$, in-game text $T_g$

- User $u$, with play history $\mathcal{H}_u$ (games played in last 7 days)

- Initial recommendation list $\mathcal{R}_u$ (up to 250 games)
}

\Output{Re-ranked top 30 recommendation list $\mathcal{R}'_u$}

\BlankLine

\textbf{Game Profile Generation:}

\ForEach{game $g \in \mathcal{G}$}{
    Extract in-game text $T_g$\;
    \If{length of $T_g$ exceeds LLM's max token limit}{
        Apply random sampling to reduce length of $T_g$\;
    }
    Generate game profile $P_g$ using LLM with prompt, based on $T_g$\;
}

\BlankLine
\textbf{LLM-Based Reranking:}

\textbf{User Profile Generation:}

Retrieve user's play history $\mathcal{H}_u$\;

Obtain game profiles $\{P_h \mid h \in \mathcal{H}_u\}$\;

Generate user profile $P_u$ using LLM with prompt, based on $\{P_h \mid h \in \mathcal{H}_u\}$\;

\BlankLine
\textbf{Personalized Reranking Strategy Generation:}

Generate personalized reranking strategy $S_u$ using LLM with prompt, based on $P_u$\;

\BlankLine
\textbf{Reranking the Top 30 List:}

Extract top 30 games $\mathcal{R}_u^{30}$ from initial recommendation list $\mathcal{R}_u$\;

Obtain game profiles $\{P_g \mid g \in \mathcal{R}_u^{30}\}$\;

Re-rank $\mathcal{R}_u^{30}$ to obtain $\mathcal{R}'_u$, based on alignment between $P_g$ and $P_u$, guided by $S_u$\;
\end{algorithm}

\section{Proposed Method}
In this section, we present a two-part pipeline designed to integrate content-driven insights into Roblox’s recommendation system. The pipeline comprises (1) Game Profile Generation, which involves extracting and structuring in-game text into detailed game profiles, and (2) LLM-Based Reranker, where these profiles are leveraged to re-rank existing recommendations and validate their effectiveness in enhancing relevance and personalization. Algorithm~\ref{alg:reranking} summarizes the overall process of our proposed method.

\subsection{Game Profile Generation}
The first component of our method is Game Profile Generation, which focuses on creating high-quality, structured profiles from raw in-game text. Given the diversity of games on Roblox and the inconsistency in developer-provided descriptions, we approach this by analyzing text that appears naturally within gameplay, as developers often guide players through instructions and cues during play. Formally, let $\mathcal{G}$ denote the set of games. For each game $g \in \mathcal{G}$, we extract the in-game text $T_g$. The main steps in this process are as follows:



\subsubsection{In-Game Text Extraction}
We extract all in-game text elements encountered by players, including instructions, background descriptions, button prompts, and other guidance, denoted as $T_g$ for each game $g$. This text provides valuable insights into the game’s genre, objectives, themes, mechanics, and language. To enable holistic analysis, we aggregate all in-game text into a single file per game, serving as input for the LLM. If $T_g$ exceeds the LLM’s token limit, we apply random sampling to reduce its length while preserving diverse content. Instead of manually filtering noisy text, we use carefully designed prompts to instruct the LLM to focus on relevant game aspects and ignore irrelevant content. This approach ensures extraction of key information, such as gameplay objectives, genre, and mechanics, without extensive preprocessing. Figure~\ref{fig:in_game_text} illustrates various types of in-game text that inform our understanding, including gameplay instructions, background context, action prompts, and language cues. This process generates structured game profiles to support enhanced recommendations

\subsubsection{Game Profile Generation via LLMs}
After preparing the in-game text file $T_g$ for each game $g$, we employ a specifically crafted prompt for the LLM to generate a structured game profile $P_g$, focusing on essential attributes while filtering out irrelevant information. The prompt directs the LLM to produce a concise summary that highlights the game’s main theme, storyline, primary objectives, and core mechanics. This summary helps the recommendation system grasp the game’s core characteristics, enabling more personalized matches with user preferences. Additionally, the LLM identifies the game genre—such as “obby” (obstacle course), “simulator,” “adventure,” or “role-playing”—which further aids in categorizing the game according to its play style. The prompt also guides the LLM to determine the target audience, considering factors such as age appropriateness (e.g., kids, teens, all ages) and gameplay appeal (e.g., casual or competitive players), ensuring that recommendations align with demographic interests.

Beyond the general overview, the prompt enables the LLM to extract key features, such as multiplayer modes, customization options, in-game purchases, and unique controls, which distinguish the game and enhance its engagement potential. It also directs the model to note any additional content, including seasonal updates, exclusive items, or special events, which contribute to the game’s dynamic appeal. If the game language is specified as “NONE,” the prompt instructs the LLM to determine the language based on the in-game text, enhancing accessibility for players by allowing language-based recommendations. Finally, the LLM assesses the game’s scale, considering aspects like game world size, level count, and gameplay duration, to provide a sense of the game’s scope and depth, which can influence player engagement. This well-structured prompt enables the LLM to output a JSON-formatted profile $P_g$ encompassing all vital attributes, essential for robust, content-driven recommendations. For the full prompt details and structure, refer to the Appendix~\ref{sec:game_profile_generation}.

\subsection{LLM-Based Reranker}
Building on the generated game profiles, the LLM-Based Reranker assesses their effectiveness by re-evaluating the initial recommendations through a content-driven, personalized process. Inspired by the Chain-of-Thought (CoT) reasoning strategy~\cite{wei2022chain}, the reranker uses carefully crafted prompts to generate a reranking strategy tailored to user preferences. By incorporating content-rich game profiles, it demonstrates how in-game text understanding enhances recommendation quality. The reranker focuses on the top 30 recommendations from the initial ranking list $\mathcal{R}_u$, which contains up to 250 games in Roblox’s "Recommended For You" sort. This focus is driven by several key factors: (1) User interaction data indicates that players predominantly engage with the top 30 games, making this subset the most relevant for optimizing user experience. (2) Improving the top 30 rankings has the highest potential impact on user satisfaction, as it directly influences the most visible and frequently accessed recommendations. (3) Reranking a smaller subset reduces computational overhead, enabling a balance between effectiveness and efficiency in the reranking process. The LLM-Based Reranker operates in three sequential steps, designed to adapt recommendations to user preferences. These include user profile generation, personalized strategy creation, and the application of this strategy to the top 30 games. Detailed prompts used for these steps are provided in the Appendix~\ref{sec:user_profile}, and Appendix~\ref{sec:game_reranking}.

\subsubsection{User Profile Generation}
The first step in the reranking process is to generate a user profile $P_u$ that encapsulates individual preferences based on the user’s recent activity. Using the generated game profiles as contextual data, we analyze the user’s play history from the past seven days, denoted as $\mathcal{H}_u$. Each game ID in this history is converted into its corresponding game profile $P_g$, producing a sequence of structured game attributes that reflect the user’s recent interactions and preferences. To ensure the validity of in-game text understanding, this process relies solely on play history and excludes any direct game IDs, which lack descriptive features. We use a carefully crafted LLM prompt to analyze this sequence of game profiles and generate a comprehensive user profile $P_u$. This prompt instructs the LLM to summarize the user’s preferences across multiple dimensions, such as game genres, themes, mechanics, and gameplay styles. For example, it considers the types of games the user played most frequently, the diversity in their preferences, and any dominant patterns in their gaming behavior.

\subsubsection{Personalized Reranking Strategy Generation}
Once the user profile $P_u$ is created, the next step is to generate a personalized reranking strategy $S_u$ tailored to the user’s preferences. This strategy is crafted by feeding the user profile into the LLM with a specially designed prompt that instructs the model to identify and prioritize attributes most relevant to the user. The prompt, inspired by the CoT reasoning approach~\cite{wei2022chain}, systematically guides the LLM to consider key aspects such as preferred game genres, gameplay mechanics, unique features, and thematic elements. The reranking strategy $S_u$ acts as a blueprint for the subsequent reranking process. For example, if a user’s profile indicates a strong preference for adventure games with exploration mechanics, $S_u$ will prioritize these attributes when re-evaluating the recommendations. The strategy explicitly avoids referencing specific game IDs, as they lack meaningful descriptive features, and instead focuses on actionable insights derived from the game profiles. 

\subsubsection{Reranking the Top 30 List}
The final step applies the personalized strategy $S_u$ to the top 30 games in the initial recommendation list, denoted as $\mathcal{R}_u^{30}$. By focusing on this concise subset, the reranker targets the most impactful portion of the ranking, where user engagement is typically highest. The LLM uses the guidelines from $S_u$ to re-evaluate and reorder these recommendations based on their alignment with the user profile $P_u$. This process leverages the structured attributes in each game profile to determine relevance and prioritization. For instance, games matching the user’s preferred genres or featuring gameplay mechanics identified in $S_u$ are ranked higher. The refined recommendation list, $\mathcal{R}'_u$, represents a personalized, content-driven ordering designed to maximize user engagement and satisfaction. By concentrating on the top 30 games, this approach balances computational efficiency with the practical needs of a user-focused recommendation system.
\begin{table*}[ht]
\setlength{\tabcolsep}{4pt} 
\renewcommand{\arraystretch}{0.9} 
\small 
\centering
\begin{tabular}{|l|c|c|c|c|c|c|c|c|c|}
\hline
\textbf{Percentile Range} & \textbf{Samples} & \textbf{Users} & \textbf{Games}  & \textbf{Sessions} & \textbf{Total Play (s)} & \textbf{Avg Play/User (s)} & \textbf{Avg User Play Length} \\
\hline
0-30\_top10 & 300 & 300 & 6,462 & 467 & 939,394 & 2,011.55 & 3.84 \\
0-30\_top20 & 300 & 300 & 6,793 & 563 & 854,987 & 1,518.63 & 3.96 \\
0-30\_top30 & 300 & 300 & 7,581 & 652 & 680,219 & 1,043.28 & 3.90 \\ \hline
30-70\_top10 & 300 & 300 & 6,550 & 524 & 684,518 & 1,306.33 & 14.00 \\
30-70\_top20 & 300 & 300 & 7,160 & 664 & 660,680 & 995.00 & 14.09 \\
30-70\_top30 & 300 & 300 & 7,243 & 651 & 598,343 & 919.11 & 14.36 \\ \hline
70-100\_top10 & 300 & 300 & 6,288 & 508 & 651,330 & 1,282.15 & 45.00 \\
70-100\_top20 & 300 & 300 & 6,871 & 598 & 678,746 & 1,135.03 & 43.80 \\
70-100\_top30 & 300 & 300 & 7,329 & 675 & 666,574 & 987.52 & 44.19 \\
\hline
\end{tabular}
\caption{Summary of Data Statistics Used in the Experimental Setup.}
\label{tab:data_statistics}
\vskip -0.2in
\end{table*}

\section{Experiments}

\subsection{Experimental Setup}

\subsubsection{Objective}
The primary objective of this experiment is to verify the effectiveness of the game profiles generated through in-game text understanding. To evaluate this, we designed an LLM-Based Reranker that utilizes only the generated game profiles and the user’s recent play history from the past seven days. By reranking recommendations based on these profiles, we aim to determine whether incorporating content-driven information enhances recommendation quality. Additionally, we engaged human annotators to assess the accuracy and relevance of the game profiles, verifying if they correctly represent the game content and align with gameplay experiences.

\subsection{Data Collection}
For this study, we extracted real training data from the Roblox platform, dividing it into nine sub-datasets to account for variations in user behavior and list length. We created these sub-datasets along two dimensions: the length of user play history and the length of the ranking list. Specifically, we used the (0-30, 30-70, 70-100) percentiles to segment user play history lengths, capturing different levels of user engagement and diversity in game interactions. Additionally, we limited the ranking list to (top10, top20, top30) games for reranking in certain scenarios. Overall, the dataset comprises 2,700 unique users and 18,941 unique games, providing a diverse sample for assessing the performance of the LLM-Based Reranker and game profiles. Table~\ref{tab:data_statistics} presents key statistics for each sub-dataset, including user and game counts as well as playtime details, providing an overview of the dataset’s characteristics and user engagement patterns.

\subsubsection{Environment}
All experiments were conducted using GPT-4o as our LLM model, selected for its ability to handle complex language understanding tasks and deliver accurate content-based interpretations. This model was used for both game profile generation and reranking, ensuring consistency across all steps of the recommendation pipeline.

\subsection{Baseline and Comparative Models}

\subsubsection{Baseline Model}
The baseline model used in this study is the existing Roblox ranking model. This model ranks games primarily based on ID-based features and user behavior data, without incorporating any content-derived attributes, such as game title, description, or thematic elements. As a traditional recommendation system, it relies on collaborative filtering methods, utilizing user and game IDs alongside sparse behavior data like play frequency. This model serves as a foundational comparison point, as it does not leverage in-game text understanding or content-driven game profiles.

\subsubsection{Comparative Models}
To evaluate the incremental impact of content-driven profiles and personalized reranking strategies, we include several comparative models:
\textbf{(1) Game Title-Based Reranking}: This model reranks the recommendation list solely based on the game title. By using only this minimal text feature, we can evaluate the basic effect of content-based reranking when limited to title information, giving a partial view of the game’s identity. \textbf{(2) Game Title and Description-Based Reranking}: This model expands on the previous approach by incorporating both game title and description for reranking. With additional descriptive context, this model provides a better understanding of the game content but still lacks the depth of the structured game profiles generated through in-game text understanding. \textbf{(3) LLM-Based Reranker without Personalized Strategy}: In this model, the LLM-Based Reranker uses the full, generated game profiles (including elements such as genre, mechanics, and thematic details) but does not apply a user-specific personalized strategy. Instead, it ranks games based on a generalized relevance derived from game profiles alone. This model helps us assess the quality and effectiveness of the generated game profiles without any individual user customization. \textbf{(4) LLM-Based Reranker with Personalized Strategy (Proposed Model)}: This is the full implementation of our proposed LLM-Based Reranker, which not only leverages content-rich game profiles but also incorporates a personalized reranking strategy based on recent user play history. By generating a custom reranking guideline for each user, this model aims to maximize recommendation relevance by aligning it closely with each user’s unique preferences and past interactions.

\begin{table*}[ht]
\centering
\renewcommand{\arraystretch}{1.2} 
\begin{tabular}{|c|c|c|c|c|c|c|c|}
\hline
\textbf{Metric} & \textbf{Percentile Range} & \textbf{Baseline} & \textbf{Title-based} & \textbf{Title+Desc} & \textbf{LLM w/o Pers.} & \textbf{Proposed} & \textbf{Improvement (\%)} \\
\hline
\multirow{3}{*}{NDCG@10} & 0-30    & \underline{0.159175} & 0.146172 & 0.143475 & 0.158127 & \textbf{0.162390} & 2.02 \\
& 30-70   & 0.133364 & 0.120783 & 0.121080 & \underline{0.134588} & \textbf{0.138708} & 3.06 \\
& 70-100  & \underline{0.151939} & 0.139790 & 0.141529 & 0.144993 & \textbf{0.165178} & 8.71 \\
& Total Avg. & \underline{0.148159} & 0.135581 & 0.135361 & 0.145902 & \textbf{0.155426} & 4.90 \\
\hline
\multirow{3}{*}{NDCG@20} & 0-30    & \underline{0.230406} & 0.200728 & 0.208124 & 0.229596 & \textbf{0.245126} & 6.39 \\
& 30-70   & \textbf{0.236136} & 0.210309 & 0.201233 & 0.211539 & \underline{0.213699} & -9.48 \\
& 70-100  & 0.224884 & 0.207583 & 0.196265 & \underline{0.230626} & \textbf{0.246512} & 6.89 \\
& Total Avg. & \underline{0.230475} & 0.206207 & 0.201874 & 0.223920 & \textbf{0.235112} & 2.01 \\
\hline
\multirow{3}{*}{NDCG@30} & 0-30    & \underline{0.309881} & 0.274296 & 0.280023 & 0.296377 & \textbf{0.314573} & 1.51 \\
& 30-70   & \textbf{0.297410} & 0.276287 & 0.278835 & 0.290023 & \underline{0.296514} & -0.30 \\
& 70-100  & \underline{0.306287} & 0.265184 & 0.261518 & 0.284544 & \textbf{0.316334} & 3.28 \\
& Total Avg. & \underline{0.304526} & 0.271922 & 0.273459 & 0.290315 & \textbf{0.309140} & 1.51 \\
\hline
\end{tabular}
\caption{Comparison of NDCG Engagement Across Different Models and Percentile Ranges. This table shows the average NDCG Engagement values, calculated over five runs to mitigate LLM instability, for the Baseline model and various reranking strategies: Title-based Reranking, Title+Desc Reranking, LLM-Based Reranker without Personalized Strategy, and the Proposed LLM-Based Reranker with Personalized Strategy. Performance is evaluated across different percentile ranges (0-30, 30-70, 70-100) for NDCG cutoffs: NDCG@10, NDCG@20, and NDCG@30. Bolded values represent the best performance, while underlined values indicate the second-best performance for each range.}
\label{tab:ndcg_engagement_comparison}
\vskip -0.1in
\end{table*}

\subsubsection{Purpose of Comparison}
The primary purpose of this comparative analysis is to evaluate the quality and effectiveness of the generated game profiles and their impact on recommendation performance. By comparing these models, we aim to measure how well each approach reflects user interests and enhances recommendation relevance. Additionally, this analysis explores the potential for applying the LLM-Based Reranker for more personalized ranking in the future. By isolating the effects of content-driven profiles and personalized strategies, we gain insight into the value of in-game text understanding and its scalability for adaptive, user-specific recommendations on Roblox.

\subsection{Evaluation Metrics}
\subsubsection{NDCG Engagement}
To evaluate the effectiveness of the LLM-Based Reranker, we use NDCG Engagement, a variation of the traditional Normalized Discounted Cumulative Gain (NDCG) metric that measures the relevance of recommendations based on user engagement. Unlike standard NDCG, where each relevant item has a binary relevance score of 1, NDCG Engagement assigns relevance scores according to the user’s playtime within the 7 days following their initial interaction with the recommended game. This approach allows us to capture deeper insights into how well the recommendations align with user interests, as increased playtime indicates a higher level of engagement and satisfaction. The formula for NDCG Engagement is as follows:
\[
\text{NDCG}_{\text{Engagement}} = \frac{1}{Z} \sum_{i=1}^{p} \frac{2^{\text{rel}(i)} - 1}{\log_2(i + 1)},
\]
where \( p \) is the position in the ranking list, \( \text{rel}(i) \) is the relevance score based on user playtime for the game at position \( i \), \( Z \) is a normalization factor ensuring that the NDCG score is between 0 and 1, calculated based on the ideal ordering of items by relevance. For clarity, all mentions of NDCG in this paper refer to NDCG Engagement unless otherwise specified.

\subsection{Results and Analysis}
\begin{figure}[t]
\begin{center}
\centerline{\includegraphics[width=\linewidth]{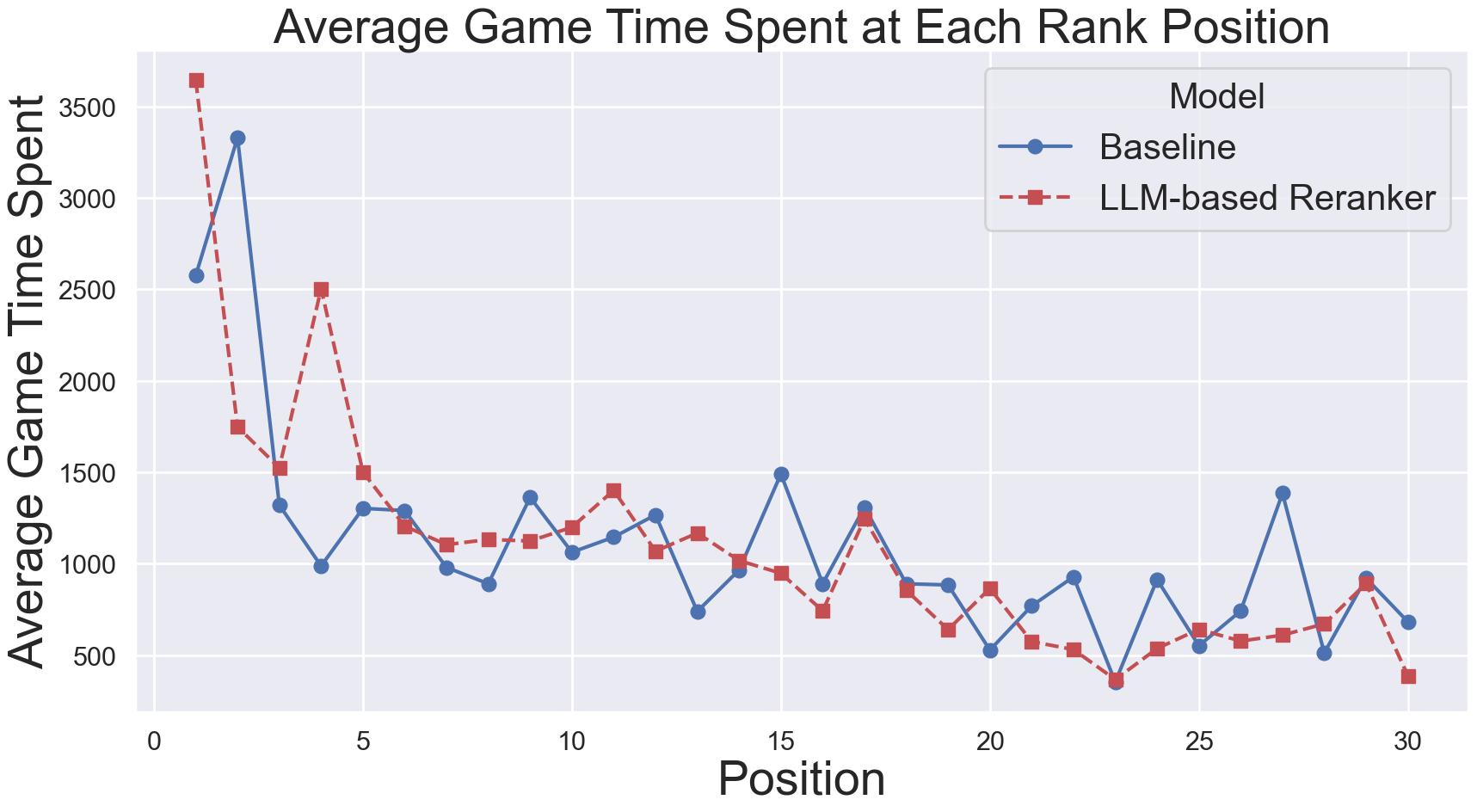}}

\caption{Average Game Time Spent at Each Rank Position. The LLM-based reranker prioritizes games that users are more likely to engage with, as shown by the higher average time spent on games ranked at top positions.}
\label{fig:user_preference}
\end{center}
\vskip -0.1in
\end{figure}

\subsubsection{Quantitative Results}
In this section, we analyze the NDCG Engagement scores for different reranking models across various percentile ranges and NDCG cutoffs (NDCG@10, NDCG@20, and NDCG@30), as shown in Table~\ref{tab:ndcg_engagement_comparison}. To ensure reliability, all experiments were run five times, and the average results are reported. The quantitative analysis highlights the performance of the Baseline model, Title-based Reranking, Title+Desc Reranking, LLM-Based Reranker without Personalized Strategy, and the Proposed LLM-Based Reranker with Personalized Strategy (proposed model).
\begin{itemize}[leftmargin=*]
    \item \textbf{Overall Improvement of the Proposed Model:} The Proposed LLM-Based Reranker with Personalized Strategy consistently demonstrates superior performance compared to the Baseline model across most percentile ranges and NDCG cutoffs. This is evident in the Improvement (\%) column, where the Proposed model achieves notable gains, especially in lower percentile ranges. The overall average improvement for the Proposed model is highest in NDCG@10 (4.90\%), followed by NDCG@20 (2.01\%), and NDCG@30 (1.51\%). These results suggest that the personalized reranking strategy is particularly effective at optimizing the top ranks, which are critical in enhancing user engagement. The substantial improvement at the higher positions (NDCG@10) emphasizes the value of personalization in capturing user interest early in the recommendation list.
    \item \textbf{Performance of Title-Based and Title+Desc Models:} Both the Title-based Reranking and Title+Desc Reranking models show a decrease in performance compared to the Baseline. While the Title+Desc model leverages both game title and description text, these features often contain noise in Roblox. Game titles and descriptions may be inconsistent or unrelated to the actual game content, as they are often created by a wide range of developers with varying levels of professionalism. Furthermore, the LLM may lack sufficient global knowledge about many Roblox games, particularly new and less popular ones, limiting its ability to interpret these text features accurately. This underlines the importance of reliable content-based features and demonstrates the limitations of using unstructured, noisy text in the absence of comprehensive domain knowledge.
    \item \textbf{LLM-Based Reranker without Personalized Strategy:} The LLM-Based Reranker without Personalized Strategy performs comparably to the Baseline model but does not outperform the Proposed model. This outcome highlights the crucial role of personalization in enhancing recommendation relevance. While the content-based reranker benefits from in-game text understanding, the absence of user-specific personalization limits its effectiveness. These findings indicate that merely incorporating content-based features is insufficient; personalized adaptation to user preferences based on in-game text understanding is essential for optimal performance.
    \item \textbf{Challenges with the 30-70 Percentile Range:} Across the 30-70 percentile range, we observe a performance decrease for most text-based models. This drop can be attributed to the diverse and less focused nature of user play histories in this range. Unlike users in the lower percentile range, who often have narrow interests, or high-engagement users, who exhibit consistent preferences, mid-engagement users tend to explore a broader variety of games without clear patterns. For example, a user might play simulation, obby, adventure, tycoon, and competition games over the past week, but their preferences may not consistently favor one genre, such as simulation, over another, like obby. Without a discernible trend in user preferences, the reranker struggles to capture specific interests accurately, especially when relying solely on in-game text features without leveraging statistical insights or behavioral patterns. This highlights the need for methods capable of resolving ambiguities in diverse play histories and creating a clearer representation of user interests for mid-engagement users.
\end{itemize}

\subsubsection{User Engagement Analysis}
Figure~\ref{fig:user_preference} illustrates the average time users spent on games at each ranking position for both the Baseline model and the LLM-based reranker. Higher values represent greater user engagement, suggesting increased interest in games at those positions. The LLM-based reranker, indicated by the dashed red line, generally places games with higher user engagement towards the top of the ranking list, particularly in the highest-ranked positions, as evidenced by the elevated average time spent. This demonstrates the LLM-based reranker’s effectiveness in surfacing games that align with user preferences, as compared to the Baseline model.

\subsubsection{Ablation Study: Impact of Different LLM Models}

To evaluate the impact of different LLM models on the performance of our proposed pipeline, we conducted an ablation study using several LLMs: Meta-Llama-3.1-8B-Instruct, Meta-Llama-3.1-70B-Instruct, GPT-4o-mini, and GPT-4-turbo. We compared each model's performance against GPT-4o, which serves as the baseline, using NDCG@30 as the evaluation metric. The results reveal significant variability in the effectiveness of the LLMs. Meta-Llama-3.1-8B-Instruct was unable to follow the instructions in our prompt and, therefore, could not produce usable results. Among the remaining models, all exhibited performance degradation compared to GPT-4o. Specifically, Meta-Llama-3.1-70B-Instruct, GPT-4o-mini, and GPT-4-turbo showed relative decreases of -3.889\%, -0.832\%, and -2.269\% in NDCG@30, respectively.

\subsubsection{Case Study: Effectiveness of LLM-Based Personalized Ranking Strategy}
\begin{figure}[t]
\begin{center}
\centerline{\includegraphics[width=\linewidth]{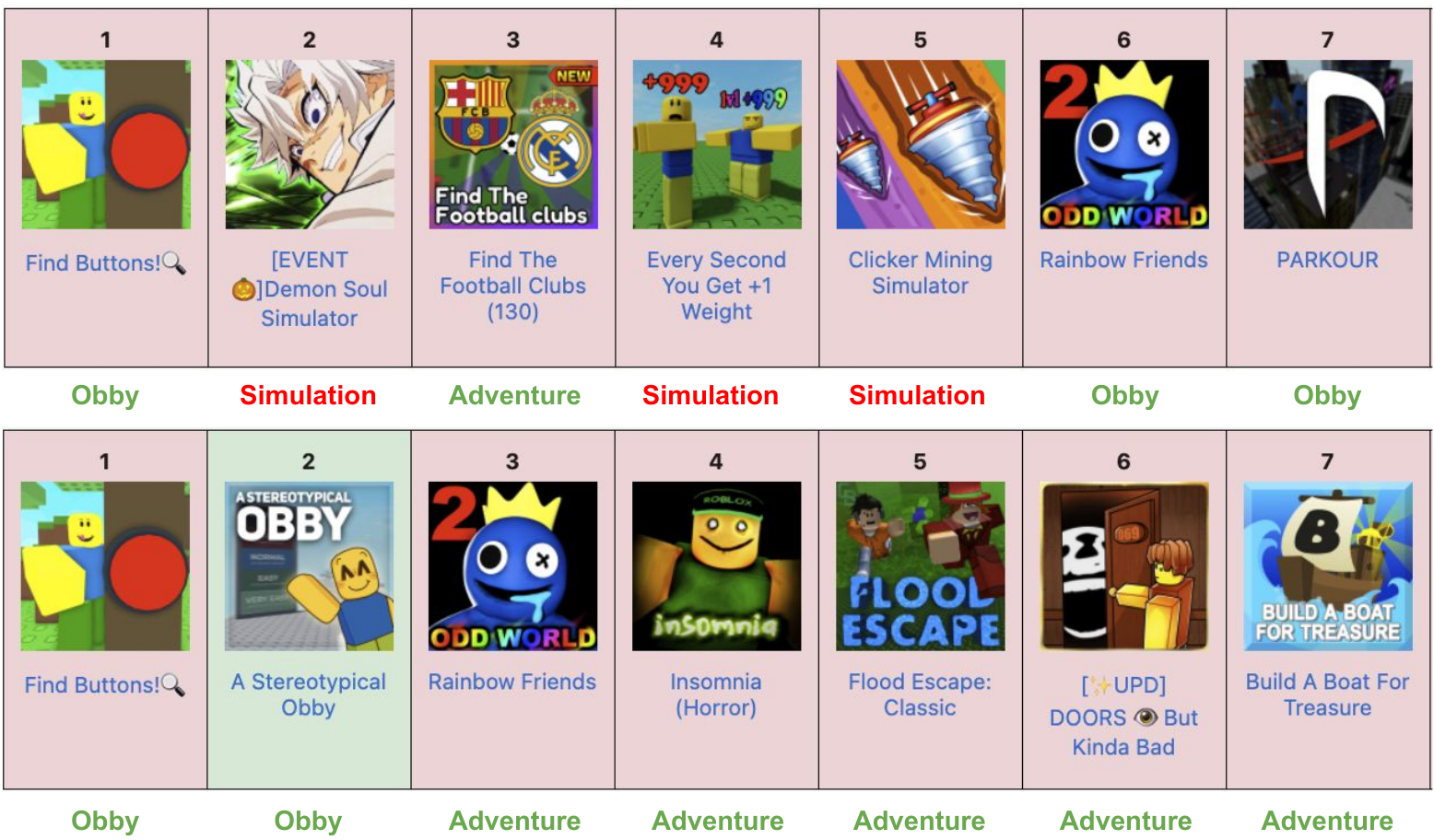}}

\caption{The first row shows the original ranking list, and the second row displays the reranked results. The LLM-based reranker successfully prioritizes high-relevance games, such as adventure and obby genres, moving them to top positions based on user preferences.}
\label{fig:case_study_ranking_list}
\end{center}
\vskip -0.3in
\end{figure}
To better understand the impact of in-game text understanding and personalized reranking, we conducted a case study on the effectiveness of the LLM-based reranker in following a personalized ranking strategy. This study demonstrates the reranker’s ability to prioritize games based on specific user preferences, highlighting its potential for delivering highly relevant recommendations. Below is an example of the ranking strategy generated by the LLM:
\begin{tcolorbox}[colback=gray!5, colframe=gray!80, title=LLM Ranking Strategy, label=strategy]
\textbf{Based on the user's preferences, the ranking logic will prioritize games as follows:}

1. \textbf{Top Priority}:
   \begin{itemize}
      \item \textcolor{red}{\textbf{Adventure}} and \textcolor{red}{\textbf{obby}} games with extensive gameplay, multiple levels, and interactive features.
      \item Games that include role-playing elements, such as character morphing and customization.
   \end{itemize}

2. \textbf{High Priority}:
   \begin{itemize}
      \item \textcolor{red}{\textbf{Adventure}} games with a focus on exploration, puzzle-solving, and dynamic activities.
      \item Games with multiplayer modes, special abilities, and rewards systems.
   \end{itemize}

3. \textbf{Medium Priority...}
\end{tcolorbox}

In this case study (see Figure~\ref{fig:case_study_ranking_list}), we demonstrate the effectiveness of our proposed LLM-based reranking model. The first row shows the original ranking list, while the second row displays the reranked results generated by our model. The LLM follows the personalized ranking strategy closely, prioritizing relevant games according to the user’s preferences as shown in Box~\ref{strategy}. For example, adventure and obby games, which were marked as high-priority, are moved to the top positions. Notably, our reranker successfully elevates highly relevant games from lower positions (such as moving a game from position 20 to position 2), showcasing its ability to refine recommendations effectively.

\subsubsection{Case Study: Enhancing Game Understanding with In-Game Text Analysis} 
\begin{figure}[t]
\begin{center}
\centerline{\includegraphics[width=\linewidth]{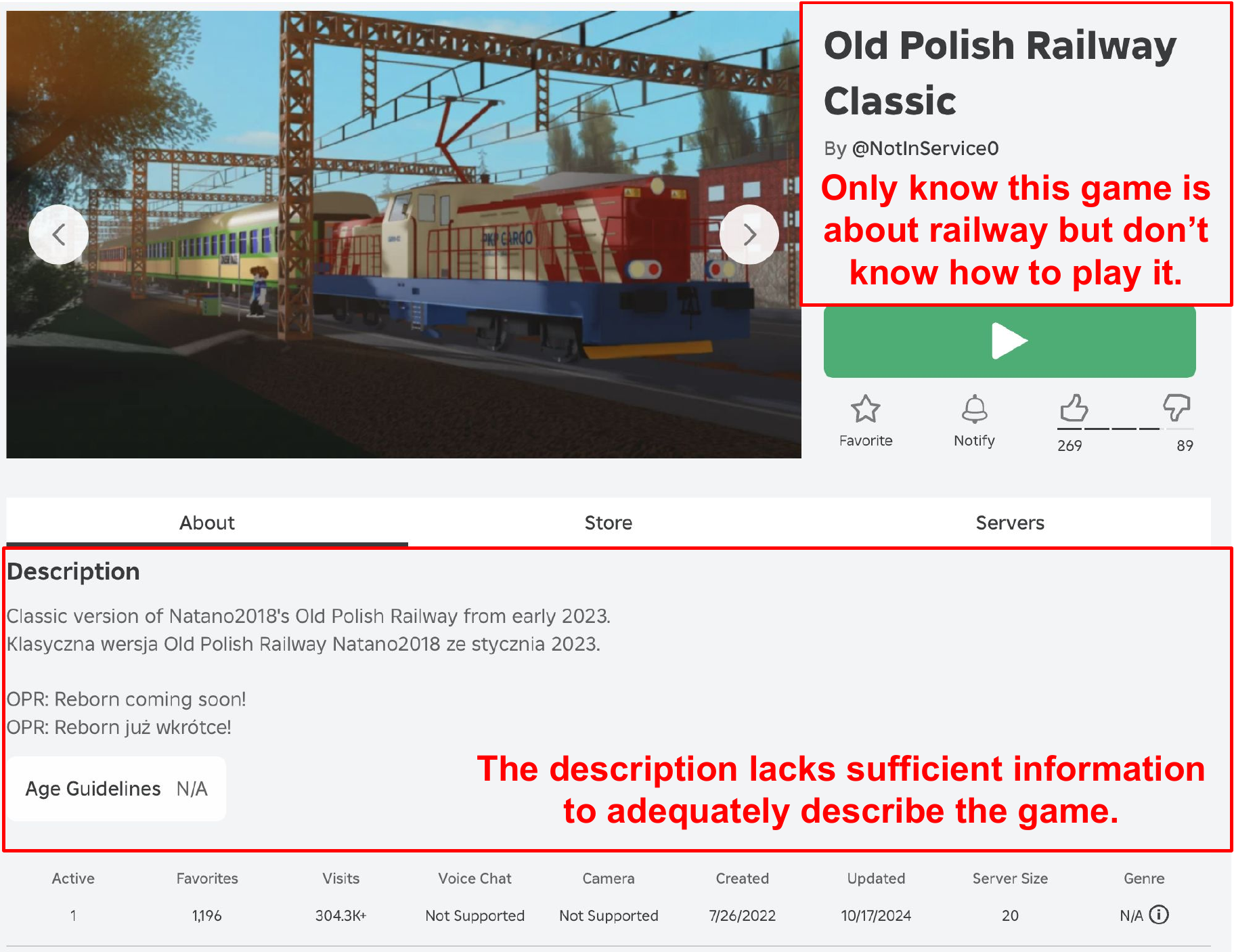}}

\caption{Example of Limited Game Information in Old Polish Railway Classic. The developer-provided title and description lack sufficient detail about gameplay, objectives, or genre, making it difficult for players to understand the game.}
\label{fig:case_study_polish_railway}
\end{center}
\vskip -0.34in
\end{figure}

In this case study (see Figure~\ref{fig:case_study_polish_railway}), we examine a Roblox game titled Old Polish Railway Classic. Based solely on the developer-provided title and description, it’s challenging to understand the gameplay, objectives, or even the specific genre of the game. The title indicates it’s related to railways, but there’s insufficient information about how to play or what players can expect. However, by leveraging our in-game text understanding model, we generated a comprehensive game profile that accurately captures the game’s content and objectives. This profile shown in Box~\ref{gamebox} provides players with a clear description of the gameplay experience, bridging the gaps left by the developer’s brief description.

\begin{tcolorbox}[colback=gray!5, colframe=gray!80, title=Game Profile about Old Polish Railway Classic, label=gamebox]
This Roblox game appears to be a railway simulator set in Poland, featuring
various Polish cities and stations. Players can manage and operate different aspects of a
railway system, including electrifying tracks, repairing platforms and switches, and handling freight cars. The game includes significant contributions from different developers and has a detailed focus on Polish railway operations.", \textcolor{red}{"game\_genre"}: "simulator",
\textcolor{red}{"suitabl\_for"}: "teens, railway enthusiasts, simulation fans", \textcolor{red}{"features"}: "multiplayer modes, detailed railway management, Polish cities and stations, electrification of tracks, repair and maintenance tasks, various control options, in-game updates and minor changes, special abilities like toggling ABS and TCS", \textcolor{red}{"includes"}: "seasonal updates, special events, exclusive items, significant contributions from developers, thank-you notes to contributors", \textcolor{red}{"game\_language"}: "Polish", \textcolor{red}{"game\_scale"}: "The game has a large scale with multiple cities and stations, detailed management tasks, and continuous updates, indicating an extensive and immersive gameplay experience.
\end{tcolorbox}

\section{Conclusion}

\subsection{Summary}
In this paper, we proposed a novel approach to improving game recommendations on Roblox by leveraging in-game text understanding and large language models (LLMs). Our method consists of two stages: Game Profile Generation, which extracts and structures raw in-game text into meaningful profiles, and an LLM-Based Reranker, which validates the effectiveness of these profiles through personalized reranking. It successfully demonstrates the value of in-game text understanding in enhancing recommendation relevance and aligning results more closely with user interests. Experimental results show that incorporating content-driven insights can improve recommendation quality, providing a strong foundation for future improvements in game recommendation systems.

\subsection{Limitations}
Despite its effectiveness, our method faces several limitations. First, the game features and taxonomies generated by LLMs remain somewhat vague and lack fine-grained distinctions. For example, while Roblox hosts a variety of games broadly categorized as simulation, adventure, or obby, user preferences often require more granular profiling to accurately capture nuanced interests. Second, the reranker, in its current form, relies solely on the LLM's general knowledge and lacks the ability to incorporate dataset-specific popularity and statistical features. This limitation arises because the LLM is not fine-tuned on the Roblox dataset, preventing it from fully leveraging platform-specific trends and user behavior patterns. Lastly, while the dynamic nature of Roblox introduces challenges in profiling new games, this can largely be addressed by profiling games only after they surpass a certain popularity or usage threshold, minimizing computational overhead. Together, these challenges highlight the need for improvements in granularity, dataset-specific adaptation, and profiling strategies.

\subsection{Future Work}
To address these limitations, future work should focus on several key areas. First, enhancing the game profile generation process to produce finer-grained and more detailed representations of user preferences and game attributes will be critical for improving personalization. Second, fine-tuning LLMs on Roblox-specific data is an essential next step to better capture platform-specific statistics and popularity trends. This fine-tuning will enable the reranker to incorporate dataset-specific insights, such as the relative popularity of games and temporal user behavior patterns, alongside general content-based understanding. Additionally, integrating multimodal data, such as game visuals, audio cues, and user interaction logs, will provide a richer context for recommendations. Future developments should also explore personalized sort generation. By leveraging the personalized reranking strategy, it would be possible to create user-specific ranking lists with tailored sort names that align with individual preferences. For example, users with a strong interest in adventure games could receive a customized "Top Adventures for You" sort, enhancing engagement through more relatable and intuitive categorizations.


\bibliographystyle{ACM-Reference-Format}
\bibliography{references}
\appendix
\section{LLM Prompt for Game Profile Generation}
\label{sec:game_profile_generation}

\begin{lstlisting}
Given a Roblox game, we have access only to the in-game text features.
These features are provided in the following list format: %s.
The game's language is: %s. If the language is specified as "NONE," please
analyze the text to determine the game's language.

Context:
1. In-game text features are the text elements displayed to users while they play the game. (WHAT)
2. These text features provide crucial information to help users understand and navigate the game. (WHY)
3. Understanding these text features is essential for users to play the game effectively. (PURPOSE)
4. The in-game text features can be noisy and may contain irrelevant information. Please focus only on the relevant information and omit any irrelevant details. (NOTE)

Task:
1. Generate a summary for the game. This summary is vital for the recommender
system to better understand the game.
2. The summary should be concise, informative, and a few sentences long. It
will help in understanding user preferences and recommending games accordingly.
3. The summary MUST be in JSON format, directly readable by json.loads(). The
JSON should have the following structure, where each key represents an
attribute of the game and the value is the corresponding attribute's value:
{
    "game_about": "Provide a concise and informative description of the Roblox
    game. Include the main theme or storyline, primary objectives, core gameplay
    mechanics, unique features, and target audience. This should give a clear
    overview of what the game is about and what players can expect.",
    "game_genre": "Specify the genre of the Roblox game. Examples include obby
    (obstacle course), tycoon, role-playing, simulator, adventure, etc. This helps
    categorize the game and gives an idea of the type of gameplay involved.",
    "suitable_for": "Indicate the target audience for the Roblox game. This
    could be based on age group (e.g., kids, teens, all ages) or other demographic
    factors (e.g., casual players, competitive players). This helps in
    understanding who the game is designed for.",
    "features": "List the key features of the Roblox game. These could include
    multiplayer modes, character customization, in-game purchases, special
    abilities, unique controls, etc. This highlights what makes the game
    interesting and engaging.",
    "includes": "Mention any additional content or elements included in the
    Roblox game. This could be special events, seasonal updates, exclusive items,
    etc. This provides information on the extra content available to players.",
    "game_language": "Specify the language of the Roblox game. If the language
    is 'NONE', analyze the in-game text to determine the language. This helps in
    understanding the linguistic accessibility of the game.",
    "game_scale": "Describe the scale of the Roblox game. This could refer to
    the size of the game world, the number of levels or stages, the length of the
    gameplay, etc. This gives an idea of the game's scope and depth."
}
\end{lstlisting}

\section{LLM Prompt for User Profile and Ranking Strategy Generation}
\label{sec:user_profile}
\begin{lstlisting}
Given the user play history in the past 7 days, please write a personalized ranking strategy for the user for future ranking usage, you can consider below attributes but not limit of them:
1. What type of games the user has played in the past 7 days?
2. What type of games the user played most frequently?
3. Analyze the user's preference based on the game genres.
4. In the ranking strategy, we do not need to mention the game ID that user has played, since game ID doesnot reflect any game features.

Below is the user play history in the past 7 days, each game is represnted by a unique id with the game profile information.
{user_play_history_str}
\end{lstlisting}

\section{LLM Prompt for Game Reranking}
\label{sec:game_reranking}
\begin{lstlisting}
Given the user's personalized ranking strategy, please rank the following games based on the user's preference.
You can use the following game profile information to rank the games.
The output format MUST be top {ranking_length} game_id list in the order of the ranking WITHOUT any other information.
Here is the user's personalized ranking strategy:
{user_profile}

Here is the game profile information for the games to be ranked:
<Candidate Game Info Start>
{ranking_results_str}
<Candidate Game Info End>
\end{lstlisting}

\end{document}